\documentclass[prl,twocolumn]{revtex4}
\usepackage{epsfig}
\usepackage{bm}

\begin{document}

\title{Hamiltonian fluid dynamics and distributed chaos}

\author{A. Bershadskii}

\affiliation{
ICAR, P.O. Box 31155, Jerusalem 91000, Israel
}

\begin{abstract}
It is shown that  distributed chaos with spontaneously broken {\it time} translational symmetry (homogeneity) has a stretched exponential frequency spectrum $E(f) \propto \exp-(f/f_0)^{1/2}$. Good agreement has been established with a laboratory experimental data obtained at large values of Rayleigh number $Ra \sim 3\cdot 10^{14}$ in thermal convection. Applications to geophysical fluid dynamics (temperature dynamics for large cities, the North Atlantic Oscillation index and the Pacific/North American pattern) have been considered.  

\end{abstract}

\maketitle

\section{Distributed chaos}

  Many dynamical systems with chaotic behaviour have the exponential power spectra \cite{oh}-\cite{fm}
$$
E(f) \propto \exp -(f/f_c)      \eqno{(1)}
$$
where $f_c = const$ is some characteristic frequency.

%%%%%%% FIGURE 1  %%%%%%%%%%%%%%%%%%
\begin{figure} \vspace{-0.6cm}\centering
\epsfig{width=.45\textwidth,file=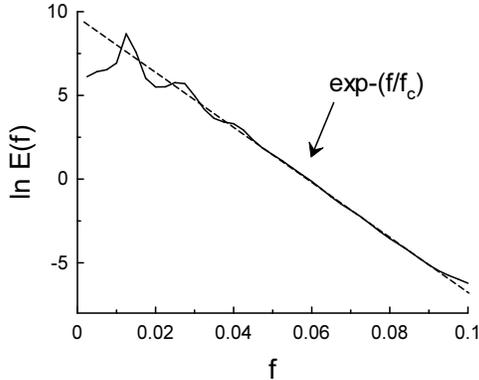} \vspace{-4.5cm}
\caption{Logarithm of power spectrum for $z$-component of the Eq. (2) against frequency $f$. The dashed straight line corresponds to Eq. (1).}
\end{figure}
%%%%%%%%%%%%%%%%%%%%%%%%%%%%%%%%%%%

 The seminal Lorenz system is a good example:
$$
\frac{dx}{dt} = \sigma (y - x),~~      
\frac{dy}{dt} = r x - y - x z, ~~
\frac{dz}{dt} = x y - b z      \eqno{(2)}          
$$
This is a very simplified model for Rayleigh-Benard (thermal) convection in a layer of fluid, cooled from above and heated from below. The parameters $b = 8/3,~ r = 28.0, ~\sigma=10.0$ provide a chaotic solution \cite{eck}. Fig. 1 shows power spectrum of $z$-component (in the semi-logarithmic scales). The spectrum was computed using the maximum entropy method, which provides an optimal resolution for comparatively short data sets \cite{oh}. The dashed straight line corresponds to Eq. (1).\\

  A weighted superposition of the exponentials Eq. (1):
$$
E(f) \propto \int_0^{\infty} P(f_c)~ \exp -(f/f_c)~ df_c   \eqno{(3)}
$$
where $P(f_c )$ is a probability distribution of $f_c$, can be used for the more complex cases of distributed chaos. 
%%%%%%% FIGURE 2  %%%%%%%%%%%%%%%%%%
\begin{figure} \vspace{-0.7cm}\centering
\epsfig{width=.45\textwidth,file=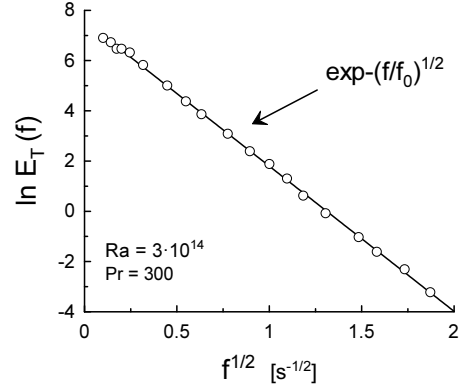} \vspace{-5cm}
\caption{Power spectrum of temperature measured at the cell {\it center} (the data taken from the Ref. \cite{as}). The solid straight line indicates the stretched exponential decay Eq. (6).}
\end{figure}
%%%%%%%%%%%%%%%%%%%%%%%%%%%%%%%%%%%

  By the Noether's theorem the energy conservation law is related to the time translational symmetry (invariance) \cite{ll},\cite{she}. When the time translational symmetry  is spontaneously broken the action $I$ can be used as an adiabatic invariant \cite{suz}. Then for this case a relation between characteristic velocity $v_c$ and the characteristic frequency $f_c$ can be found using the dimensional considerations:
$$    
v_c \propto I^{1/2} f_c^{1/2}  \eqno{(4)}
$$
If the characteristic velocity is normally (Gaussian) distributed, then $f_c$ has the chi-squared ($\chi^{2}$) distribution
$$
P(f_c) \propto f_c^{-1/2} \exp-(f_c/4f_0)  \eqno{(5)}
$$
where $f_0$ is certain constant. 

   Substituting the distribution Eq. (5) into integral Eq. (3) we obtain
$$
E(f) \propto \exp-(f/f_0)^{1/2}  \eqno{(6)}
$$
About Hamiltonian approach to hydrodynamics see, for instance, Refs. \cite{eyi}-\cite{sal} and references therein.

\section{Thermal convection}
  
  The above discussed Lorenz system is a very simplified model for thermal convection (see also Ref. \cite{tg},\cite{gl} for  Hamiltonian low-order energy conserving models of Rayleigh-Benard convection). 
Let us consider real (experimental) strong thermal convection. Figure 2 shows power spectrum of temperature measured at an upright cylinder cell's {\it center} for the Prandtl number $Pr=300$ and very large Rayleigh number $Ra=3\cdot 10^{14}$ \cite{as}. The solid straight line is drawn in the Fig. 2 in order to indicate the stretched exponential spectrum Eq. (6) corresponding to the distributed chaos with spontaneously broken translational symmetry. Similar spectrum was observed for the first time in experiment reported in Ref. \cite{wu}. In this situation the inertial (scaling) range was suppressed completely by buoyancy \cite{as}. Actually the Eq. (6) covers the entire spectrum (rather wide range) as one can see in figure 3 (in the log-log scales). It should be noted that thermal convection at the high values of the Rayleigh number is usually considered as strong turbulence and chaos was previously taken into account at the large scale thermal winds only \cite{sbn},\cite{b1}. 
%%%%%%% FIGURE 3  %%%%%%%%%%%%%%%%%%
\begin{figure} \vspace{-1cm}\centering
\epsfig{width=.45\textwidth,file=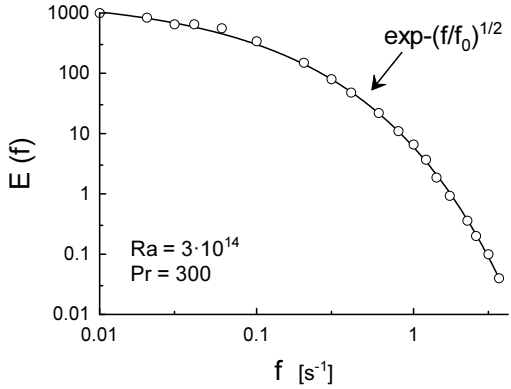} \vspace{-4.7cm}
\caption{As in Fig. 2 but in the log-log scales. The solid curve indicates the stretched exponential decay Eq. (6).}
\end{figure}
%%%%%%%%%%%%%%%%%%%%%%%%%%%%%%%%%%%

\section{Applications to geophysical fluid dynamics}

%%%%%%% FIGURE 4 %%%%%%%%%%%%%%%%%%
\begin{figure} \vspace{-1.5cm}\centering
\epsfig{width=.45\textwidth,file=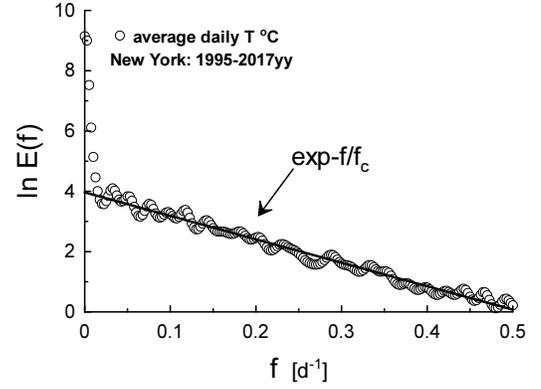} \vspace{-4.2cm}
\caption{Power spectrum of the daily average temperature fluctuations for New York (1995-2017yy) \cite{day}.}
\end{figure}
%%%%%%%%%%%%%%%%%%%%%%%%%%%%%%%%%%%

%%%%%%% FIGURE 5 %%%%%%%%%%%%%%%%%%
\begin{figure} \vspace{-0.4cm}\centering
\epsfig{width=.45\textwidth,file=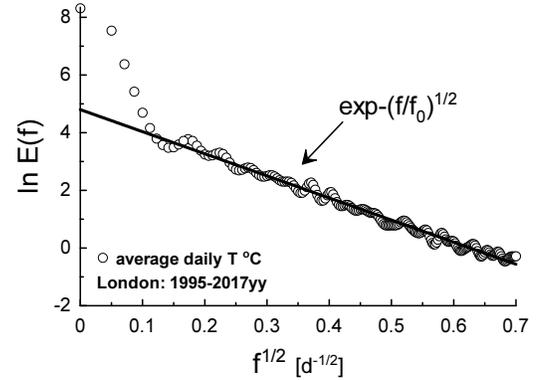} \vspace{-4cm}
\caption{Power spectrum of the daily average temperature fluctuations for London (1995-2017yy) \cite{day}.}
\end{figure}
%%%%%%%%%%%%%%%%%%%%%%%%%%%%%%%%%%%
%%%%%%% FIGURE 6 %%%%%%%%%%%%%%%%%%
\begin{figure} \vspace{-0.4cm}\centering
\epsfig{width=.45\textwidth,file=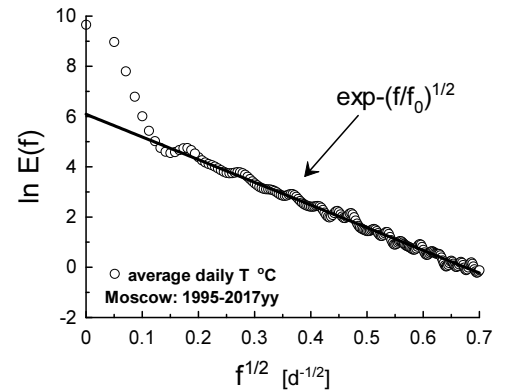} \vspace{-4.5cm}
\caption{As in Fig. 5 but for Moscow.}
\end{figure}
%%%%%%%%%%%%%%%%%%%%%%%%%%%%%%%%%%%
  The Lorenz system Eq. (2) was developed as a model of atmospheric thermal convection \cite{lor}. On the other hand, most of the important theoretical models in geophysical fluid dynamics are Hamiltonian \cite{she},\cite{gl},\cite{sh},\cite{gl2}. Therefore, one can expect that Eq. (6) can be applied to climate dynamics as well. 
%%%%%%% FIGURE 7 %%%%%%%%%%%%%%%%%%
\begin{figure} \vspace{-1.5cm}\centering
\epsfig{width=.45\textwidth,file=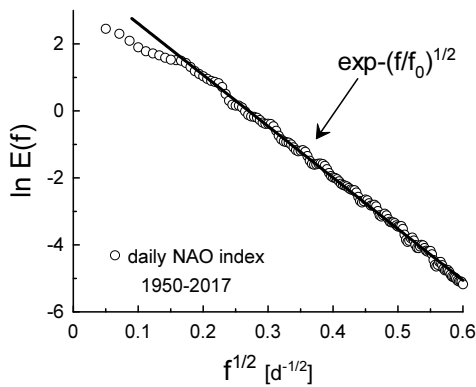} \vspace{-4cm}
\caption{Power spectrum for the daily NAO index 1950-2017yy \cite{NAO}. }
\end{figure}
%%%%%%%%%%%%%%%%%%%%%%%%%%%%%%%%%%% 
\subsection{Temperature dynamics for New York, London and Moscow}

Figure 4 shows power spectrum of average daily temperature fluctuations for New York City for period 1995-2017 years (the data were taken from the site Ref. \cite{day}). The spectrum was computed by the maximum entropy method, that gives an optimal resolution for comparatively short time series \cite{oh}. The solid straight line is drawn in the Fig. 4 in order to indicate exponential spectrum Eq. (1) ($T_c =1/f_c \simeq 8$d).

   Figure 5 shows power spectrum of average daily temperature fluctuations for London for period 1995-2017 years (the data were taken from the site Ref. \cite{day}). The solid straight line is drawn in the Fig. 5 in order to indicate stretched exponential spectrum Eq. (6) corresponding to the distributed chaos with spontaneously broken translational symmetry ($T_0 =1/f_0 \simeq 59$d). Figure 6 shows power spectrum of average daily temperature fluctuations for Moscow for period 1995-2017 years (the data were taken from the site Ref. \cite{day}). The solid straight line is drawn in the Fig. 6 in order to indicate stretched exponential spectrum Eq. (6) corresponding to the distributed chaos with spontaneously broken translational symmetry ($T_0 =1/f_0 \simeq 81$d). 
   
  Corresponding autocorrelation functions for London and Moscow (the distributed chaos) have prominent exponential tails.
  
\subsection{North Atlantic Oscillation and the Pacific/North American indexes} 
 
     Figure 7 shows a spectrum computed using the daily NAO index 1950-2017yy \cite{NAO} (NAO - North Atlantic Oscillation, one of the most prominent and influential patterns in the North Atlantic: see, for a comprehensive review \cite{hu}). The solid straight line is drawn in the Fig. 7 in order to indicate the stretched exponential spectrum Eq. (6) corresponding to the distributed chaos with spontaneously broken translational symmetry. Parameter $T_0 =1/f_0 \simeq 235$d in this case.
     
     Let us now use the simplest possible Haar wavelet regression for a detrending of the daily NAO index. In order to kill or keep the wavelet coefficients we used the VisuShrink soft thresholding function. Figure 8 shows a spectrum computed using the detrended daily NAO index 1950-2017yy. The solid straight line is drawn in the Fig. 8 in order to indicate the stretched exponential spectrum Eq. (6) corresponding to the distributed chaos with spontaneously broken translational symmetry ($T_0=1/f_0 = 160$d). It should be noted that a deep detrending of the daily NAO index (with symmlet regression of a high order) reveal a pure exponential background spectrum. A spectral peak, corresponding to period $T \simeq 44$d, is revealed by the detrending (indicated by an arrow in the Fig. 8). The deep detrending indicates a peak with $T \simeq 40$d. The approximately 40-day oscillations are also known for the Northern Hemisphere extratropics \cite{mag},\cite{mar}. 
%%%%%%% FIGURE 8 %%%%%%%%%%%%%%%%%%
\begin{figure} \vspace{-1.2cm}\centering
\epsfig{width=.45\textwidth,file=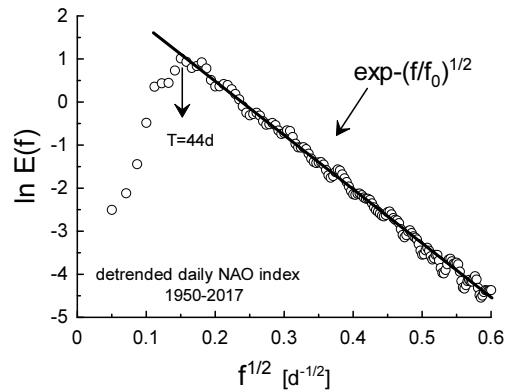} \vspace{-4.3cm}
\caption{Power spectrum for the detrended daily NAO index 1950-2017yy \cite{NAO}.}
\end{figure}
%%%%%%%%%%%%%%%%%%%%%%%%%%%%%%%%%%%
%%%%%%% FIGURE 9 %%%%%%%%%%%%%%%%%%
\begin{figure} \vspace{-1.5cm}\centering
\epsfig{width=.45\textwidth,file=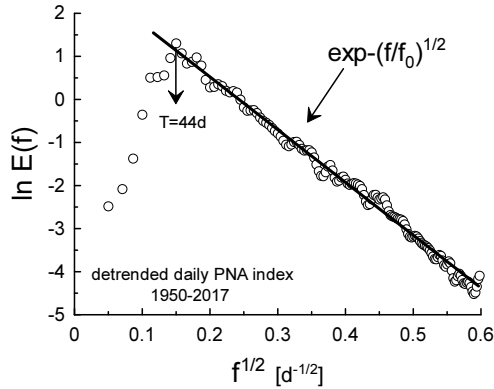} \vspace{-4.5cm}
\caption{As in Fig. 8 but for PNA.}
\end{figure}
%%%%%%%%%%%%%%%%%%%%%%%%%%%%%%%%%%%
%%%%%%% FIGURE 10 %%%%%%%%%%%%%%%%%%
\begin{figure} \vspace{-0.5cm}\centering
\epsfig{width=.45\textwidth,file=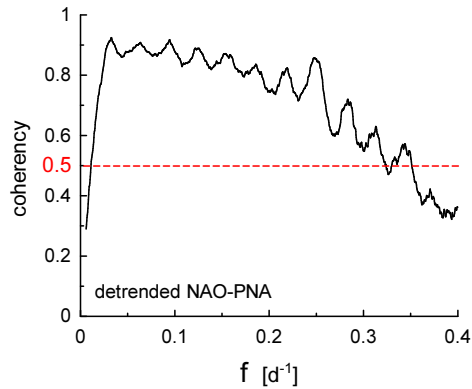} \vspace{-4.4cm}
\caption{The coherency for detrended NAO and PNA indexes.}
\end{figure}
%%%%%%%%%%%%%%%%%%%%%%%%%%%%%%%%%%%   
     Figure 9 shows a spectrum computed using the Haar wavelet detrended daily PNA index 1950-2017yy (see for the data set \cite{PNA}). PNA - Pacific/North American pattern is one of the most prominent and influential low-frequency modes in the Northern Hemisphere extratropics. The solid straight line is drawn in the Fig. 9 in order to indicate the stretched exponential spectrum Eq. (6) corresponding to the distributed chaos with spontaneously broken translational symmetry ($T_0 =1/f_0 \simeq 146$d). The PAN spectrum is very similar to that shown for NAO index in the Fig. 8.\\
     
    Let us also compute the cross spectrum of the detrended NAO and PNA indexes. For two processes $x_1(t)$ and $x_2(t)$ the cross spectrum 
$E_{1,2}(f)$ is defined as:
$$
E_{1,2}(f)= \frac{\sum_{\tau} \langle x_1(t)x_2(t-\tau) \rangle 
\exp (-i2\pi f \tau)}{2\pi \sqrt{E_1(f)E_2(f)}} \eqno{(7)}
$$
where the expectation value is defined by bracket $\langle... \rangle$. One can use a decomposition into the coherency $C_{1,2}(f)$ and the phase spectrum $\phi_{1,2} (f)$:
$$
E_{1,2}(f)= C_{1,2}(f) e^{-i \phi_{1,2} (f)}  \eqno{(8)}
$$
The coherency is ranging from $C_{1,2}(f)=1$, i.e. perfect linear relationship, to $C_{1,2}(f)=0$, i.e. no linear relationship between $x_1(t)$ and $x_2(t)$ at a frequency $f$. The coherency $\simeq 0.5$ is usually considered as a boundary between high and low coherency. Figure 10 shows the coherency computed for detrended NAO and PNA indexes using the Blackman-Tukey method
(the maximum entropy method can be applied for single time series only).

\section{Acknowledgement}

I thank G.L. Eyink for sending to me the Ref. \cite{eyi} and explanations, and J. Hurrell for sharing the NAO data. I acknowledge using the data provided by University of Dayton - Environmental Protection Agency (USA) and NOAA/ National Weather Service (USA).


\begin{thebibliography}{99}

 
\bibitem{oh} N. Ohtomo, K. Tokiwano, Y. Tanaka et. al., J. Phys. Soc.
Jpn. {\bf 64} 1104 (1995).
\bibitem{sig} D.E. Sigeti, Phys. Rev. E, {\bf 52}, 2443 (1995).
\bibitem{f} J. D. Farmer, Physica D, {\bf 4}, 366 (1982).
\bibitem{fm}U. Frisch and R. Morf, Phys. Rev., {\bf 23}, 2673 (1981).
\bibitem{eck} R. E. Ecke, Chaos {\bf 25}, 097605 (2015).
\bibitem{ll} L.D. Landau and E.M. Lifshitz, Mechanics (Pergamon Press 1969).
\bibitem{she} T.G. Shepherd, Advances in Geophysics, {\bf 32},  287 (1990).
\bibitem{suz} R.Z. Sagdeev, D.A. Usikov, G.M. Zaslavsky, Nonlinear Physics: from the 
Pendulum to Turbulence and Chaos (Harwood, New York, 1988).
\bibitem{eyi} G.L. Eyink, Physica D {\bf 239}, 1236 (2010).
\bibitem{mor} P.J. Morrison, Hamiltonian Fluid Mechanics, Encyclopedia of Mathematical Physics. {\bf 2}, 593 (Elsevier, Amsterdam, 2006).
\bibitem{yz}  V. Yakhot, V. Zakharov, Physica D {\bf 64}, 379 (1993).
\bibitem{sal} R. Salmon, Annual Review of Fluid Mechanics. {\bf 20}, 225, (1988).
\bibitem{tg} C. Tong and A. Gluhovsky, Phys. Rev. E {\bf 65}, 046306 (2002).
\bibitem{gl} A. Gluhovsky, Nonlinear Processes in Geophysics, {\bf 13}, 125 (2006).
\bibitem{as} S. Ashkenazi and V. Steinberg, Phys. Rev. Lett. {\bf 83}, 3641 (1999).
\bibitem{wu} X-Z. Wu, L. Kadanoff, A. Libchaber, and M. Sano, Phys. Rev. Lett. {\bf 64}, 2140 (1990).
\bibitem{sbn} K. R. Sreenivasan, A. Bershadskii and J. J. Niemela, Phys Rev E {\bf 65}, 056306 (2002).
\bibitem{b1} A. Bershadskii, Chaos {\bf 20}, 043124 (2010).
\bibitem{lor} E.N. Lorenz,  J. Atm. Sci. {\bf 20} 130 (1963).
\bibitem{sh} T.G.Shepherd, Encyclopedia of Atmospheric Sciences, J. R. Holton et al., Eds.,
929 (Academic Press, 2003).
\bibitem{gl2} A. Gluhovsky, and K. Grady, Chaos, {\bf 26}, 023119 (2016).
\bibitem{day} http://academic.udayton.edu/kissock/http/Weather/
\bibitem{NAO} http://www.cpc.ncep.noaa.gov/products/precip/CWlink/
pna/nao.shtml
\bibitem{hu} J.W. Hurrell et al., in ”The North Atlantic Oscillation: Climatic Significance and Environmental Impact”, Geophysical Monograph {\bf 134}, p.1, American Geophysical Union (2003).
\bibitem{o2} T. Ogden, Essential Wavelets for Statistical Applications and Data Analysis 
(Birkhauser, Basel, 1997).
\bibitem{mag} V. Magana, J. Geophys. Res., {\bf 98}, 10441 (1993).
\bibitem{mar} S.L. Marcus, M. Ghil, and J.O. Dickey, J. Atmos. Sci., {\bf 51}, 1431 (1994).
\bibitem{PNA} http://www.cpc.ncep.noaa.gov/products/precip/CWlink/
pna/pna.shtml






\end{thebibliography}
\end{document}